\documentclass[showpacs,showkeys]{revtex4}
\usepackage{amsmath}
\usepackage{amssymb}
\usepackage{graphicx}

\begin{document}
\title{Wormhole solutions supported by interacting dark matter and dark energy
}
\author{
Vladimir Folomeev$^{1,2,4}$
\footnote{vfolomeev@mail.ru}
and Vladimir Dzhunushaliev$^{1,2,3,4}$
\footnote{v.dzhunushaliev@gmail.com}
}
\affiliation{
$^1$Institut f\"ur Physik, Universit\"at Oldenburg, Postfach 2503,
D-26111 Oldenburg, Germany
\\
$^2$Institute of Physicotechnical Problems and Material Science, National Academy of Sciences
of the Kyrgyz Republic, 265~a, Chui Street, Bishkek 720071,  Kyrgyz Republic
\\
$^3$Department of Theoretical and Nuclear Physics, Kazakh National University, Almaty 050040, Kazakhstan
\\
$^4$IETP, Kazakh National University, Almaty 050040, Kazakhstan
}

\begin{abstract}
We show that  the presence of a nonminimal interaction between dark matter and dark energy may lead to a
violation of the null energy condition and to the formation of a configuration with nontrivial topology (a wormhole).
In this it is assumed that
both dark matter and dark energy satisfy the null energy condition, a violation of which
takes place only in the inner high-density regions of the configuration.
This is achieved by assuming that, in a high-density environment,
a nonminimal coupling function changes its sign
in comparison with the case where dark matter and dark energy
have relatively low densities which are typical for a cosmological background.
For this case, we find regular static, spherically symmetric solutions describing
wormholes supported by
dark matter nonminimally coupled to dark energy in the form of a quintessence scalar field.
\end{abstract}

\pacs{04.40.Dg,  04.40.--b, 97.10.Cv}
\keywords{Interacting dark matter and dark energy; wormholes; nontrivial topology}
\maketitle

\section{Introduction}

Since  its creation,  Einstein's general relativity has proved to be a powerful tool
used in considering various systems for which relativistic gravitational effects are significant.
It has been used successfully both in describing  the cosmological evolution of the early and present Universe and
in modeling various strongly gravitating compact objects. In the latter case, perhaps the most exotic objects predicted by
general relativity are black holes and wormholes.
It is commonly believed that such objects might be formed in regions of strong gravitational fields.
The principal feature of any wormhole is the presence of
a throat, defined
as a two-dimensional hypersurface of minimal  area.
This throat can connect either two distant regions of our Universe or  even different universes,
and far from the throat, spacetime may either be flat or have a curved geometry.

A necessary condition to construct a throat is the presence of matter violating
the so-called null energy condition~(NEC),
$T_{\mu \nu} k^\mu k^\nu \geq 0$, where $k^\mu$ is any null vector, and $T_{\mu \nu}$ is the energy-momentum tensor of  matter.
In general,  $T_{\mu \nu}$ can contain different types of matter, for instance, fluids and/or various  fundamental fields.
In the simplest case of a  perfect isotropic fluid having only one radial pressure, $p_r$,
and the energy density $\varepsilon$,
the NEC can be presented in the form $(\varepsilon+p_r)\geq 0$.
Then, if such a fluid is described by a simple equation of state in the
form  $p_r=w\varepsilon$, a violation of the NEC becomes possible when the equation-of-state parameter $w<-1$.

This form of matter is called exotic matter. It has been used both in describing the present accelerated expansion
of the Universe \cite{sahni:2004,Copeland:2006wr,AmenTsu2010,Bamba:2012cp} and in constructing such exotic objects as wormholes
\cite{Bronnikov:1973fh,Ellis:1973yv,Kodama,lxli,ArmendarizPicon:2002km,Sushkov:2002ef,Visser:2003yf,Lemos:2003jb,lobo,sushkov,
Dzhunushaliev:2007cs,kuhfittig,Dzhunushaliev:2010bv}. A quantity of greater interest are models
 of a traversable Lorentzian wormhole,
suggested by Morris and Thorne  \cite{Thorne:1988} (for a review, see the book by Visser \cite{Visser}).
The traversability assumes that matter and radiation
can travel freely through the wormhole which can lead to some interesting observable consequences \cite{Kardashev:2006nj}.

There are several possibilities to introduce exotic matter. This can be a consideration of systems with the above-mentioned
linear equation of state $p_r=w\varepsilon$ with  $w<-1$ \cite{lobo,sushkov}, or the use of
 ghost/phantom scalar fields
 \cite{Bronnikov:1973fh,Ellis:1973yv,Kodama,lxli,ArmendarizPicon:2002km,Dzhunushaliev:2007cs,Dzhunushaliev:2010bv}, or a consideration of systems
 beyond the bounds of Einstein gravity \cite{Sushkov:2002ef,BEG}. In any case, to build
a model of a static traversable wormhole in general relativity, one has to violate the
 NEC \cite{Hochberg:1997wp} which requires the use of exotic matter,  in one form or another, even if
in arbitrarily small quantities \cite{Visser:2003yf,Kardashev:2006nj}.

Here the question arises whether such exotic forms of matter can exist in the Universe.
A strong argument in favor of such a possibility
is the observed accelerated expansion of the present Universe.
The numerous attempts to give a theoretical description of the current accelerated expansion of the Universe are usually reduced
to introducing
a new form of matter, called dark energy (DE), which is believed to be responsible for such an acceleration.
Having large negative pressure, dark energy causes the Universe
to expand increasingly fast, according to a power law or exponentially.
Moreover, astronomical observations  (see, e.g., Refs.~\cite{Tonry:2003zg,Alam:2003fg}
and more recent estimates \cite{Sullivan:2011kv})
indicate the possibility, that an even more exotic form of
energy exists in the Universe, called phantom dark energy.
The presence of such an energy assumes the violation of the NEC
and results in even faster acceleration.

The amount of dark energy is about 70\% of the total energy density of the Universe. Another important component
of the present Universe is dark matter (DM) which contributes about 25\% of the total energy density, and
is clustered on scales of the order of galaxies and clusters of galaxies.
Although  the true nature of neither DM nor DE is currently known,
various ways have been suggested to model them. In the simplest case this can be the so-called $\Lambda$CDM model,
where DE is described by Einstein's $\Lambda$ term, and DM is supposed to be
a pressureless fluid (cold dark matter). However, due to the well-known cosmological constant problem~\cite{sahni:2004,Copeland:2006wr,AmenTsu2010,Bamba:2012cp},
alternative possibilities are being suggested to describe the evolution of the present Universe.
One of the approaches to address the origin of DE and DM are theories that involve
various fundamental fields~\cite{sahni:2004,Copeland:2006wr,AmenTsu2010,Bamba:2012cp}, including
cosmological models with DE and DM interacting with each other not only
gravitationally but also by a direct coupling
\cite{Damour:1990tw,Wetterich:1994bg,Amendola:1999er,Billyard:2000bh,Bartolo:1999sq,Zimdahl:2001ar,
Maccio:2003yk,Farrar:2003uw,Das:2005yj,Amendola:2006qi,Guo:2007zk,Boehmer:2008av,Bean:2008ac}
(for a review, see, e.g., Refs.~\cite{Copeland:2006wr,AmenTsu2010,Bamba:2012cp,Tsujikawa:2010sc}).
In turn, the supposed presence in the Universe of DE,  in one form or another,
and DM
provides the basis for concluding
that compact systems consisting of dark energy
\cite{de_stars}, of dark matter
\cite{dm_stars,Narain:2006kx}, or of
interacting DE and DM \cite{Brouzakis:2005cj,de_dm_conf}
might also exist.

In the present paper, we study compact, spherically symmetric configurations consisting of DE and DM
nonminimally interacting with each other.
In doing so, we model DE in the form of a usual (nonghost) quintessence-type
scalar field, and DM is described as a perfect fluid satisfying the NEC.
Here our objective will be to construct  regular static solutions
describing self-consistently a configuration with nontrivial topology, where the violation of the
NEC is achieved by an appropriate choice of the form of a nonminimal coupling between dark matter and dark energy.

The paper is organized as follows.
In Sec.~\ref{Lagr_choice} we discuss the choice of the dark matter/dark energy
interaction Lagrangian, using which we derive the general set of equations for equilibrium configurations in
Sec.~\ref{gen_eqns_wh_de_dm}.
In Sec.~\ref{statem_prob} the statement of the problem is presented, for which in Sec.~\ref{throat_cond} we find
the necessary conditions providing the existence of a throat in the system under consideration.
Using these conditions, in Sec.~\ref{num_res} we consider
an explicit example of obtaining regular static solutions with nontrivial topology with a particular choice
of a nonminimal coupling function.
 Finally, in Sec.~\ref{conclusion}  our results are summarized.

\section{Derivation of the equations for equilibrium configurations}

\subsection{The dark matter/dark energy interaction Lagrangian}
\label{Lagr_choice}

At the moment, there exists no fundamental theory that allows one to choose
a specific coupling in the dark sector. Therefore, any type of coupling
will necessarily be phenomenological, although some models may appear to be more physically motivated than others.
In modeling DE in the form of a scalar field, one can
meet in the literature different types of couplings  between the field and dark matter.
This can be the possibility, inspired by scalar-tensor theories of gravity, when one assumes the presence of an
interaction of the form $Q\, T_{(\text{DM})} \varphi_{,\nu}$, where
$T_{(\text{DM})}$ is the trace of the energy-momentum tensor
of dark matter, $\varphi$ is the cosmological scalar field,
 and $Q$ can be a constant \cite{Wetterich:1994bg,Amendola:1999er,Maccio:2003yk} or be field dependent \cite{Billyard:2000bh,Bartolo:1999sq,Amendola:2006qi}.

If one initially works within the framework of
general relativity,  the appearance of a direct coupling between the scalar field and matter is also possible  when the
mass of matter particles is assumed to be explicitly dependent on the scalar field.
This can be an exponential dependence
(for cosmological implications, see Ref.~\cite{yuk_exp}) or a linear dependence appearing as a consequence
of the presence of the Yukawa coupling. The latter type of interaction  has been repeatedly considered
 in the literature. In particular,  it was used  in studying compact objects in Refs.~\cite{Brouzakis:2005cj,Lee:1986tr,Crawford:2009gx},
 in describing structure formation in Ref.~\cite{Nusser:2004qu}, and in modeling  interacting DM and DE
 in Refs.~\cite{Farrar:2003uw,Bean:2008ac}.

An explicit dependence of the mass of particles on a scalar field implies that by going
to a description of matter in the form of a fluid its pressure and density  become  functions
depending explicitly on the scalar field (see, e.g., Ref.~\cite{Brouzakis:2005cj}). However, one can  use instead a
phenomenological possibility when a description of the interaction between the
scalar field and matter is performed assuming that  the pressure and density are not initially explicit functions of
a scalar field. Such a coupling can be expressed through an interaction  Lagrangian of the form
\begin{equation}
\label{lagr_int}
L_{\text{int}}=f(\varphi) L_m,
\end{equation}
where $L_m$ is the Lagrangian of matter (ordinary or dark) and
the coupling function $f(\varphi)$ characterizes the
interaction between $\varphi$ and the matter.
The case $f\to 1$ corresponds to the absence of a direct coupling between
matter and the scalar field, when the two sources are coupled only via gravity.

In the cosmological context,  the interaction Lagrangian \eqref{lagr_int} was used  in modeling
the present accelerated  expansion of the Universe
\cite{Beans,Das:2005yj,Koivisto:2005nr,Farajollahi:2010pk,Cannata:2010qd,Folomeev:2012sz},
in describing structure formation \cite{struc_form_f_phi},
and when considering compact configurations
\cite{cham_stars,Folomeev:2012sz}. In doing so, a choice
of the Lagrangian $L_m$ is, in general, not unique.
It can be taken as
$L_m=-\varepsilon$ \cite{Hawking1973}
or  $L_m=p$~\cite{Stanuk2}, where $\varepsilon$ and $p$ are the energy density and the  pressure of an isentropic perfect fluid.
By varying both these  matter Lagrangians with respect to a metric, one obtains the same energy-momentum tensor
of the perfect
fluid in the conventional form. However, it can be shown that, for instance, for static configurations, these Lagrangians
will give different equations for an equilibrium
configuration (see, e.g., the discussion in Ref.~\cite{Folomeev:2012sz}).
Apparently, at the present time, it is difficult to make a motivated choice among these Lagrangians
 $L_m$, or any other Lagrangians  used in the literature
(for the other possible Lagrangians and a discussion of their choice for nonminimally coupled systems
see, e.g., Ref.~\cite{Bertolami:2008ab}). For this reason, a description of various systems with nonminimal coupling of the type
 $f(\varphi) L_m$ is made, in effect,
using an {\it ad hoc} choice of  $L_m$.

\subsection{General set of equations}
\label{gen_eqns_wh_de_dm}

As pointed out in the Introduction, the objective of the present paper is to study the possibility of obtaining solutions describing static
configurations with nontrivial topology sourced by a usual (nonghost) cosmological scalar field interacting with dark matter both gravitationally and
through a direct coupling.
The general Lagrangian for such a system can be presented in the form
\begin{equation}
\label{lagr_SFDMEF}
L=-\frac{c^4}{16\pi G}R+\frac{1}{2}\partial_{l}\varphi\partial^{l}\varphi -V(\varphi)+f(\varphi) L_{\text{DM}}
 -\frac{1}{4}F_{lm}F^{lm}~.
\end{equation}
Here $\varphi$ is the real scalar field with the potential $V(\varphi)$,
$F_{lm}$ is the the electromagnetic field tensor, and 
$L_{\text{DM}}=-\varepsilon$ is the Lagrangian of dark matter, where $\varepsilon$ denotes the energy density of dark matter.
The case $f=1$ corresponds to the absence of a direct coupling between
dark matter and the scalar field. However, even in this case the two sources are still coupled via gravity.

The energy-momentum tensor can be
obtained by varying the matter part of the Lagrangian \eqref{lagr_SFDMEF} with respect to a metric,
\begin{equation}
\label{emt_cham_wh}
T_i^k=f\left[(\varepsilon+p)u_i u^k-\delta_i^k p\right]+ \partial_{i}\varphi\partial^{k}\varphi-F^l_i F^k_l
-\delta_i^k\left[\frac{1}{2}\,\partial_{l}\varphi\partial^{l}\varphi-V(\varphi)-\frac{1}{4}F_{lm}F^{lm}\right],
\end{equation}
where  $p$ is the pressure of a dark matter fluid,  and $u^i$ is the four-velocity.

In considering equilibrium wormhole-like configurations, we will use the polar Gaussian coordinates
\begin{equation}
\label{metric_wh}
ds^2=A(l) (d x^0)^2- dl^2-r^2(l)\, d\Omega^2,
\end{equation}
where $A$ and $r$ are functions of the radial coordinate $l$, the time coordinate  $x^0=c\, t$,
and $d\Omega^2$ is the metric on the unit two-sphere.
The coordinate $l$ covers the entire range $(-\infty, +\infty)$.

Here we consider only radial components of electric and magnetic fields,
$F_{01}=E_r$ and $F_{23}=-H_r$, with the following ansatz for the magnetic field:
 $F_{23}=\partial_{\theta}A_3$, where $A_3=Q_m \cos\theta$,  $Q_m$ is the magnetic
charge,  and $\theta$ is the angular coordinate on a sphere.
Then, using Maxwell's equations $\left[\sqrt{-g}F^{ik}\right]_{,k}=0$, one can find $E_r=Q_e/r^2$,
where $Q_e$ is the electric charge.

For our purposes, we use
the $(_0^0)$, $(_1^1)$, and $(_2^2)$ components of the Einstein equations, which
can be obtained by using the metric  \eqref{metric_wh} and the energy-momentum tensor \eqref{emt_cham_wh}
in the following form, respectively (hereafter we work in natural units $c=\hbar=1$):
\begin{eqnarray}
\label{Einstein-00_cham_wh}
&&-\left[2\frac{R^{\prime\prime}}{R}+\left(\frac{R^\prime}{R}\right)^2\right]+\frac{1}{R^2}=
8\pi \left[f \varepsilon+\frac{1}{2}  \phi^{\prime 2}+V(\phi)+\frac{\beta}{2}\frac{Q^2}{R^4}\right],
 \\
\label{Einstein-11_cham_wh}
&&-\frac{R^\prime}{R}\left(\frac{R^\prime}{R}+\frac{A^\prime}{A}\right)+\frac{1}{R^2}=
8\pi \left[-f p-\frac{1}{2}\phi^{\prime 2}+V(\phi)+\frac{\beta}{2}\frac{Q^2}{R^4}\right],
\\
\label{Einstein-22_cham_wh}
&&\frac{R^{\prime\prime}}{R}+\frac{1}{2}\frac{A^\prime}{A}\frac{R^\prime}{R}+
\frac{1}{2}\frac{A^{\prime\prime}}{A}-\frac{1}{4}\left(\frac{A^\prime}{A}\right)^2=
8\pi \left[f p-\frac{1}{2}\phi^{\prime 2}-V(\phi)+\frac{\beta}{2}\frac{Q^2}{R^4}\right],
\end{eqnarray}
where the prime denotes  differentiation with respect to the radial coordinate.
Here we have introduced dimensionless variables,
\begin{equation}
\label{dimless_var}
\xi=\frac{l}{L}, \quad R=\frac{r}{L}, \quad \phi=\varphi/M_{\text{Pl}}, \quad
\tilde{\varepsilon}=\frac{\varepsilon}{m^4}, \quad \tilde{p}=\frac{p}{m^4},
\quad
\tilde{V}(\phi)=\frac{V(\varphi)}{m^4}
\quad
\text{with} \quad L=\frac{M_{\text{Pl}}}{m^2},
\end{equation}
where $m$ is the mass of dark matter particles, and
$M_{\text{Pl}}$ is the Planck mass. Also, the parameter $\beta=\left(m/M_{\text{Pl}}\right)^4$ and $Q^2=Q_e^2+Q_m^2$.
For convenience, we drop the tilde in
Eqs.~\eqref{Einstein-00_cham_wh}-\eqref{Einstein-22_cham_wh} and hereafter.

Since not all of the Einstein field equations are independent because of the
conservation of energy and momentum,
$T^k_{i;k}=0$, the $i=1$ component of this equation gives
\begin{equation}
\label{conserv_2_cham_wh}
\frac{d p}{d \xi}=-(\varepsilon+p)\left(\frac{1}{2}\frac{A^\prime}{A}+\frac{1}{f}\frac{d f}{d\phi}\phi^\prime\right).
\end{equation}

Equations \eqref{Einstein-00_cham_wh}-\eqref{Einstein-22_cham_wh} and \eqref{conserv_2_cham_wh} must be supplemented by an equation
for the scalar field which follows from the Lagrangian~\eqref{lagr_SFDMEF},
$$
\frac{1}{\sqrt{-g}}\frac{\partial}{\partial x^i}\left[\sqrt{-g}g^{ik}\frac{\partial \varphi}{\partial x^k}\right]=
-\left(\frac{d V}{d \varphi}+\varepsilon \frac{d f}{d \varphi}\right),
$$
and gives
\begin{equation}
\label{sf_eq_dmls}
\phi^{\prime\prime}+\left(\frac{1}{2}\frac{A^\prime}{A}+2\frac{R^\prime}{R}\right)\phi^\prime
=
\frac{d V}{d\phi}+\varepsilon\frac{d f}{d\phi}.
\end{equation}

Thus, we have five unknown functions:
$A, R, \phi, \varepsilon$, and $p$.
Keeping in mind that $\varepsilon$ and $p$ are related by an  equation of state,
there remains only four unknown functions.
To determine these functions, we can use
 any three equations from the system \eqref{Einstein-00_cham_wh}-\eqref{Einstein-22_cham_wh},  \eqref{conserv_2_cham_wh}
together with
the scalar-field equation \eqref{sf_eq_dmls}.

\section{An explicit example}

\subsection{Statement of the problem}
\label{statem_prob}

Here we  consider
the conditions under which the system
described by the Lagrangian \eqref{lagr_SFDMEF} may permit the existence of configurations with nontrivial topology.
In doing so, we proceed from the following assumptions.
\begin{enumerate}
\itemsep=-0.2pt
\item[(i)] 
Dark matter, being clustered  in a finite region of space, has the boundary at some
$\xi=\xi_b$ where its energy density and pressure become equal to zero,
$\varepsilon(\xi_b), p(\xi_b)=0$.
\item[(ii)] At large distances ($\xi \gg \xi_b$) the coupling function $f$ tends to 1.
\item[(iii)] At large distances ($\xi \gg \xi_b$) the scalar field $\phi$ goes to some constant (cosmological) value $\phi_0$
providing the positive energy density of dark energy close to the critical energy density $\simeq 10^{-47} \text{GeV}^{4}$,
which represents the averaged cosmological energy density in the Universe today.
This in turn implies the positivity of the potential energy $V(\varphi)$.
\item[(iv)] It is assumed that both the dark matter and the dark energy
do not violate  the NEC.
\end{enumerate}

Then, taking into account that at large distances the energy density of the electric and magnetic fields is much smaller
than the scalar-field energy density (see below in Sec.~\ref{num_res}),
the resulting Lagrangian
becomes identical in form to that used in describing the cosmological evolution of the Universe. In this sense,
configurations being studied here may be thought of as embedded in a cosmological background.

To obtain a nontrivial topology in the system under consideration, it is necessary to violate the NEC in the vicinity
of the center.
Here we will consider solutions symmetric with respect to the center with the boundary conditions given by Eq.~\eqref{bound_wh}
(see below). Since, by definition, a throat
is a two-dimensional hypersurface of minimal  area, then the  areal radius
$R$  must have a minimum at the throat. Without loss of generality, we can take this throat to occur at $\xi=0$.
Then, mathematically, we need to: i) provide a minimum of the function $R(\xi)$ at the point $\xi=0$, and
ii)~ensure that $R(0)$ is real.
Taking into account that the derivative of the scalar field is equal to zero at the throat
[see Eq.~\eqref{bound_wh}] and the condition (iv),
it will be shown in Sec.~\ref{throat_cond} that
 the function   $R$ has a minimum only when  $f<0$ in the vicinity of the center.

As discussed in Sec.~\ref{Lagr_choice}, at the present time the choice of the function
 $f$ is strictly model dependent, and does not follow from any first principles.
It is usually assumed that $f$ is everywhere positive, and is chosen
in such a way as to be compatible with  current astronomical and cosmological observations.
The latter implies, in particular, that  $f$ gives an adequate description of a
nonminimal interaction between dark matter
and a scalar field  at small energy densities which are typical for cosmology and galactic halos.

On the other hand, at higher energy densities which are typical for inner parts of compact objects, the
possibility is not excluded that conditions might occur when $f$ becomes negative. In particular,
it may be due to quantum effects which might lead to  violations
of the NEC \cite{Visser:2003yf}. In any case, this important question requires special consideration,
which goes beyond the scope of this paper. Here we wish only to analyze  some general requirements for the appearance of
nontrivial topology in the system described by the Lagrangian
\eqref{lagr_SFDMEF}
with the proviso that the above conditions (i)-(iv) are satisfied.

We note here that in the absence of a nonminimal interaction the literature in the field offers
alternative ways to violate the NEC only in the vicinity of a throat. In particular, in Ref.~\cite{Bronnikov:2010hu}
the case was considered where a scalar field (which can change the sign of the ``kinetic'' term
depending on the magnitude of the field)
 becomes a ghost near a throat, remaining nonghost
everywhere else, as in our case.

Taking account of the choice $f<0$ in the vicinity of the center and the condition
(ii), the function $f$, as it moves outward from  the center of the configuration to large distances,
must necessarily go through zero at some point. Since  $f$ stands in the denominator of Eq.~\eqref{conserv_2_cham_wh},
this certainly implies that  $d f/d\phi$  also must be
equal to zero at that point. It appears that this condition cannot be realized, at least for analytic functions.
One way around this difficulty
is to choose an $f$ that crosses zero
somewhere outside the fluid where Eq.~\eqref{conserv_2_cham_wh} is already absent.

Another possibility is to choose a special
equation of state for  dark matter. In the simplest case it can be
a linear equation of state in the form
$p=w\varepsilon$, where $w=\text{const}$, which is widely used  in cosmology.
In this case Eq.~\eqref{conserv_2_cham_wh} can be integrated in the form
\begin{equation}
\label{enrg_f_A}
\varepsilon=\varepsilon_c \left(\frac{f_c}{f}\sqrt{\frac{A_c}{A}}\,\right)^{(w+1)/w}.
\end{equation}
Here the index ``{\it c}'' corresponds to central values of the variables at $\xi=0$.
Assuming that the metric function $A$ remains finite and nonzero everywhere, it is seen from this expression that to provide
the regularity of $\varepsilon$ along the radius of the system we must take $w<0$.

The most popular hypothesis in modeling dark matter
is the assumption that its equation of state corresponds to cold dust matter \cite{sahni:2004}.
Such matter is either pressureless or has a very small pressure with the equation-of-state parameter
lying in the range $-10^{-2}\lesssim w \lesssim 10^{-3}$ \cite{Muller:2004yb}.
This assumes that the pressure can be negative.
On the other hand, when modeling dark matter by a scalar field, it is possible for
$w$ to be considerably smaller than zero \cite{Bharadwaj:2003iw}. Consistent with this, in what follows  we  will
consider the possibility of obtaining a nontrivial topology in our system.

\subsection{Conditions at the throat}
\label{throat_cond}

Let us now turn to a consideration of boundary conditions for the set of equations
\eqref{Einstein-00_cham_wh}-\eqref{Einstein-22_cham_wh} and  \eqref{sf_eq_dmls}.
Since here we will seek  solutions which are symmetric with respect to the center,
 we choose  boundary conditions in the neighborhood of the center of the configuration as
\begin{equation}
\label{bound_wh}
R \simeq R_c+\frac{1}{2}R_2 \xi^2, \quad A \simeq A_c+\frac{1}{2}A_2 \xi^2, \quad
\phi \simeq \phi_c+\frac{1}{2}\phi_2 \xi^2,
\end{equation}
where the index ``{\it c}'' corresponds to the central values of the variables at $\xi=0$.
Substituting these boundary conditions and the expression for
 $\varepsilon$ from Eq.~\eqref{enrg_f_A} into Eqs.~\eqref{Einstein-00_cham_wh}-\eqref{Einstein-22_cham_wh}  and \eqref{sf_eq_dmls},
 we find the following expressions. For the throat radius,
Eq.~\eqref{Einstein-11_cham_wh} gives
\begin{equation}
\label{bound_Rc}
R_c^2=\frac{1\pm \sqrt{1-8\pi Q^2 \beta D}}{D} \quad \text{with} \quad D=16\pi(V_c-w f_c\, \varepsilon_c).
\end{equation}
From Eq.~\eqref{Einstein-00_cham_wh} we have
\begin{equation}
\label{bound_R2}
R_2=-4\pi (1+w) R_c f_c\, \varepsilon_c,
\end{equation}
and Eq.~\eqref{Einstein-22_cham_wh} yields
\begin{equation}
\label{bound_A2}
A_2=8\pi A_c\left[ (1+3w) f_c\, \varepsilon_c-2\left(V_c-\frac{\beta}{2}\frac{Q^2}{R_c^4}\right)\right].
\end{equation}
Finally, from Eq.~\eqref{sf_eq_dmls} we have
\begin{equation}
\label{bound_phi2}
\phi_2=\left(\frac{d V}{d\phi}\right)_c+\varepsilon_c\left(\frac{d f}{d\phi}\right)_c.
\end{equation}

In these expressions $V_c$ and $f_c$ refer to the central values of the corresponding functions of the scalar field.
Since here we deal only with negative values of $w$ and $f_c$, then we always have $w f_c\, \varepsilon_c>0$.
Then it is seen from Eq.~\eqref{bound_Rc} that $D$ can be negative or positive depending on the value of the
parameters appearing in $D$. In order to ensure that $R$ is a real function, we have the two following sets of
boundary conditions which can be realized at the throat:
\begin{eqnarray}
\label{cond_A}
\text{(A)}&& D>0, \quad (1-8\pi Q^2 \beta D)\geqslant 0, \quad (1\pm \sqrt{1-8\pi Q^2 \beta D})>0,\\
\text{(B)}&& D<0, \quad (1-8\pi Q^2 \beta D)\geqslant 0, \quad (1- \sqrt{1-8\pi Q^2 \beta D})<0.
\label{cond_B}
\end{eqnarray}

Both these sets are supplemented by two more conditions providing a minimum of corresponding functions:
$R_2~\geqslant~0, A_2~\geqslant~0$. Then we have from Eqs.~\eqref{bound_R2} and \eqref{bound_A2}, respectively,
\begin{equation}
\label{cond_R2_A2}
f_c\, \varepsilon_c\leqslant 0, \quad \left[(1+3w) f_c\, \varepsilon_c-2\left(V_c-\frac{\beta}{2}\frac{Q^2}{R_c^4}\right)\right]\geqslant 0.
\end{equation}
(Note here that the condition $A_2\geqslant 0$, which ensures a minimum of the function $A$,
need not be necessary in general. Examples of regular static solutions in models with scalar fields
where this condition is not satisfied can be found in Ref.~\cite{Dzhunushaliev:2007cs}.)

It can be shown  that in the absence of the electric and magnetic fields, i.e., when $Q=0$, the inequalities \eqref{cond_A}-\eqref{cond_R2_A2}
cannot be fulfilled simultaneously. That is why we will solve these inequalities simultaneously in the case when $Q\neq 0$.
Then it is possible to find a range of allowed values for the parameters of the system such that these inequalities are fulfilled.
For the case (A), there are only the two following sets of conditions:
\begin{eqnarray}
\label{cond_A1_Q_fc}
\text{(A1)}:&& Q^2=\gamma_1 \frac{R_c^2}{8\pi \beta}, \quad f_c=-\frac{\gamma_1\gamma_2}{8\pi (1+w)R_c^2 \varepsilon_c}, \\
\label{cond_A1_Rc}
&& R_c^2=\frac{(1+w)(2-\gamma_1)-2 w \gamma_1\gamma_2}{16\pi(1+w)V_c},\\
&&\text{with}\quad 1<\gamma_1<2,\quad \frac{\gamma_1^2-1}{\gamma_1^2} \leqslant \gamma_2 \leqslant \frac{2(\gamma_1-1)}{\gamma_1},
\quad -1<w<0,\nonumber\\
\label{cond_A2_Q_fc}
\text{(A2)}:&& Q^2=\gamma_1 \frac{R_c^2}{8\pi \beta},
\quad f_c=-\frac{\gamma_2}{\gamma_1}\,\,\frac{\gamma_1^2-1}{8\pi (1+w) R_c^2  \varepsilon_c}, \\
\label{cond_A2_Rc}
&& R_c^2=\frac{(1+w)(2-\gamma_1)\gamma_1-2 w (\gamma_1^2-1)\gamma_2}{16\pi(1+w)\gamma_1 V_c},\\
&&\text{with}\quad 1<\gamma_1<2,\quad 0 < \gamma_2 < 1,
\quad -1<w<0.\nonumber
\end{eqnarray}
The free parameters $\gamma_1$ and $\gamma_2$ appearing here are arbitrary and limited only in the ranges shown above.

For the case (B), the number of ranges of allowed values for the parameters is much larger.
In order to not encumber the paper, here we show only one set of conditions which will be used below in performing
numerical calculations:
\begin{eqnarray}
\label{cond_B1_Q_fc}
\text{(B1)}:&& Q^2 =-\frac{w \gamma_2 \gamma_3}{16\pi^2 (1+w)\beta V_c},
\quad f_c=\frac{w \gamma_2^2 \gamma_3}{16\pi^2 (1+w)^2 V_c R_c^4 \varepsilon_c}, \\
\label{cond_B1_Rc}
&& R_c^2=\frac{1}{16\pi V_c}\left[1-\sqrt{1+\frac{8 w \gamma_2\gamma_3[1+w(1+2\gamma_2)]}{(1+w)^2}}\,\right],\\
&&\text{with}\quad 0<\gamma_2 < 1, \quad  0 < \gamma_3 < 1,
\quad -\frac{1}{1+4\gamma_2}<w<0.\nonumber
\end{eqnarray}

From the expressions obtained above, we can find the relation between the throat radius $r_c$
and the central value of the intensity of the electric and/or magnetic fields, $H_{c}=Q_{\text{cgs}}^2/r_c^2$, in the
cgs system of units. Namely, going back to dimensional quantities, we have
\begin{eqnarray}
\label{rel_rc_Hc_A}
\text{(A1), (A2)}:&& r_c=\sqrt{\frac{\gamma_1}{8\pi G}}\frac{c^2}{H_c}, \\
\label{rel_rc_Hc_B}
\text{(B1)}:&& r_c =\alpha_B\sqrt{\frac{1}{\pi G }}\frac{c^2}{H_c}
\quad\text{with}\quad
\alpha_B=\sqrt{-\frac{w\gamma_2\gamma_3}{1+w}\left[1-\sqrt{1+\frac{8 w\gamma_2\gamma_3\left[1+w(1+2\gamma_2)\right]}{(1+w)^2}}
\,\right]^{-1}}.
\end{eqnarray}
Depending on the value of the arbitrary parameters $\gamma_2, \gamma_3$,  and also on the equation-of-state parameter
$w$, the coefficient $\alpha_B$ lies in the range
$1/2<\alpha_B<1/\sqrt{2}$. I.e., the relations between $r_c$ and $H_c$ in all cases
(A1), (A2), and (B1) remain approximately the same.
Similar relations between $r_c$ and $H_c$ have been found in Ref.~\cite{Kardashev:2006nj}
for magnetic wormholes supported by a radial magnetic field with a small amount of exotic matter.
In that case, the magnetic field is, in essence, a main component of such wormholes, since it determines their sizes and masses
(see Table 1 of Ref.~\cite{Kardashev:2006nj}).

Calculating the coefficient  $\alpha_B$ in other ranges of allowed values for the parameters of the system, one can show
that a minimum value of $\alpha_B$  remains always of the order of 1, but a maximum value may be much larger, for example,
in the case when there is a possibility that $w \to -1$. Then, for some fixed value of
$H_c$, the throat radius, and correspondingly its mass, will increase
as $w$ tends to $-1$. On the other hand,
since a minimum value of $\alpha_B$ is~$\sim {\cal O}(1)$, then it appears difficult to decrease considerably the sizes and masses
of the systems under consideration in comparison with those of Ref.~\cite{Kardashev:2006nj}.

\subsection{Numerical results}
\label{num_res}

The restrictions on the parameters of the system considered in the previous section are necessary conditions
assuring the possibility of the existence of a nontrivial topology in the system. However, they still do not guarantee
that there exist solutions satisfying the conditions (i)-(iv) from Sec.~\ref{statem_prob}.
Here we demonstrate the possibility for obtaining such solutions for definite choices of the scalar functions
$f(\phi)$ and $V(\phi)$.

Let us seek numerical solutions of Eqs.~\eqref{Einstein-00_cham_wh}, \eqref{Einstein-22_cham_wh}, and  \eqref{sf_eq_dmls}
with the boundary conditions \eqref{bound_wh} and some particular form of the functions
 $f(\phi)$ and $V(\phi)$. As already discussed above,
$f(\phi)$ is a strictly model-dependent function. Hence,  taking into account the conditions
(ii) and (iii) from Sec.~\ref{statem_prob}, we may choose it, e.g., in the following form:
\begin{equation}
\label{fun_f}
f=1+a (\phi-\phi_0)^b,
\end{equation}
where $a,b$ are free parameters and $\phi_0$ is the cosmological value of the scalar field whose magnitude depends upon
the specific form of the potential energy  $V(\phi)$.
Such a coupling function can arise as
a consequence of  a conformal transformation from the string frame into the Einstein
frame \cite{Beans}.

The free parameters $a,b$ must be such as to provide the necessary central value
$f_c\equiv f(\phi_c)$, given by the expressions
\eqref{cond_A1_Q_fc}, \eqref{cond_A2_Q_fc}, or \eqref{cond_B1_Q_fc}. Since these expressions depend directly on the central
value of the potential energy $V_c\equiv V(\phi_c)$, on which the throat size also depends directly
[see Eqs.~\eqref{cond_A1_Rc}, \eqref{cond_A2_Rc}, and \eqref{cond_B1_Rc}], then $V_c$ is, in essence,
the primary  determining parameter of the problem. Its value depends on the form of the potential energy used,
and may vary within wide limits.

As a potential energy, here we use that from Ref.~\cite{Brax:1999gp}, which can be presented in the variables
being used here as follows:
\begin{equation}
\label{fun_V}
V=M^{4}\left(M/M_{\text{Pl}}\right)^\alpha \phi^{-\alpha} \exp{(\lambda \phi^2)}.
\end{equation}
Such a function has been used in describing the present accelerated expansion of the Universe within the
framework of the so-called ``freezing'' models of quintessence \cite{AmenTsu2010,Tsujikawa:2010sc}.
This potential energy is always positive and
has a minimum at $\phi=\phi_{\text{min}}$ where the cosmological quintessence
field is eventually trapped. In cosmological applications, the two free parameters
$\alpha$ and $\lambda$ appearing here, and also the mass scale $M$, are adjusted
to provide conditions under which the tracking behavior is realized
(see, e.g., Ref.~\cite{AmenTsu2010}). For instance, using the value of $\alpha=11$ and choosing
$\lambda=4\pi$~\cite{Brax:1999gp}, a minimum of the potential \eqref{fun_V} takes place at
 $\phi_{\text{min}}\simeq 0.66$
(recall that here $\phi$ is measured in units of the Planck mass). In describing the evolution of the current Universe
using the potential \eqref{fun_V}, it is assumed that the scalar field varies slowly with time and stays
for a sufficiently long time in the vicinity of the minimum of the potential, thus providing
a sufficiently prolonged stage of accelerated expansion.
That is, it is assumed that the current value of the scalar field $\phi_0 \simeq \phi_{\text{min}}$.
 Assuming that
$V(\phi_0) \simeq 10^{-47} \text{GeV}^{4}$ -- the current averaged cosmological density in the Universe --
the mass scale $M$ is given by $M\simeq 4 \times 10^{10}~\text{GeV}$~\cite{Brax:1999gp}. In subsequent calculations we will use these
values of the parameters $\alpha, \lambda$, and $M$.

Using the above functions $f(\phi)$ and $V(\phi)$,
we solve the system of equations \eqref{Einstein-00_cham_wh}, \eqref{Einstein-22_cham_wh}, and  \eqref{sf_eq_dmls} subject to the boundary conditions
\eqref{bound_wh}. We start the numerical procedure at the center $\xi = 0$ and proceed to the point
$\xi = \xi_b$,  where the nonminimal coupling function $f$ goes to zero.
Since in the present paper we reckon the equation-of-state parameter $w$ as negative,
then, according to
Eq.~\eqref{enrg_f_A}, the energy density and the pressure of the dark matter also vanish at the point $\xi_b$.
We refer to the obtained solutions as internal solutions.
Since the wormhole configurations under consideration
are supposed to be embedded in an external, homogeneously distributed cosmological scalar field $\phi_0$, then,
to provide the smoothness of solutions along the radius,
we require the internal solutions to match external solutions obtained for the region $\xi > \xi_b$ characterized by nonzero energy densities
of the scalar and electric/magnetic fields.
Thus, for $\xi > \xi_b$, we proceed with numerical solutions
of Eqs.~\eqref{Einstein-00_cham_wh}, \eqref{Einstein-22_cham_wh}, and  \eqref{sf_eq_dmls} retaining
only the gravitational, electric/magnetic, and scalar fields while the dark matter fluid is taken to be zero.
These external solutions are interrupted
at the point where $\phi=\phi_0$, i.e., in the vicinity of a minimum of the potential \eqref{fun_V},
and we also
adjust the free parameters of the system in such a way that the gradient of the scalar field becomes equal to zero at this point.

As an example, we show the results of numerical calculations for the case (B1) [see
Eqs.~\eqref{cond_B1_Q_fc} and \eqref{cond_B1_Rc}]. Here it is convenient to choose the free parameter
 $\gamma_3$ in the form
\begin{equation}
\label{expr_g3}
\gamma_3=-\frac{4\pi V_c (1-8\pi V_c)(1+w)^2}{w \gamma_2 [1+w(1+2\gamma_2)]}.
\end{equation}
Substituting this expression in Eqs.~\eqref{cond_B1_Q_fc} and \eqref{cond_B1_Rc}, we have
\begin{equation}
\label{expr_R_c_f_c}
R_c=1, \quad f_c=-\frac{(1-8\pi V_c)\gamma_2}{4\pi [1+w(1+2\gamma_2)]\varepsilon_c},
\end{equation}
and the expression for $Q$ can be found from Eq.~\eqref{cond_B1_Q_fc}   using
$\gamma_3$ from Eq.~\eqref{expr_g3}. Then, taking some specific value of the mass of a dark matter particle  $m$,
one can obtain solutions for different values
of the equation-of-state parameter $w<0$. Another free parameter
$\gamma_2$ is then chosen so that
at $\xi \gg \xi_b$ the scalar field goes to its cosmological value
$\phi_0$ in the vicinity of a minimum
of the potential \eqref{fun_V} with the corresponding cosmological $V(\phi_0) \simeq 10^{-47}~\text{GeV}^{4}$.
Since we seek solutions for which the gradient of the scalar field at this point is
$\phi^\prime_{\phi=\phi_0}=0$, this enables the solution for the scalar field to be matched smoothly onto the
cosmological background  solution $\phi=\phi_0$.

\begin{figure}[t]
\begin{minipage}[t]{.5\linewidth}
  \begin{center}
  \includegraphics[width=7.5cm]{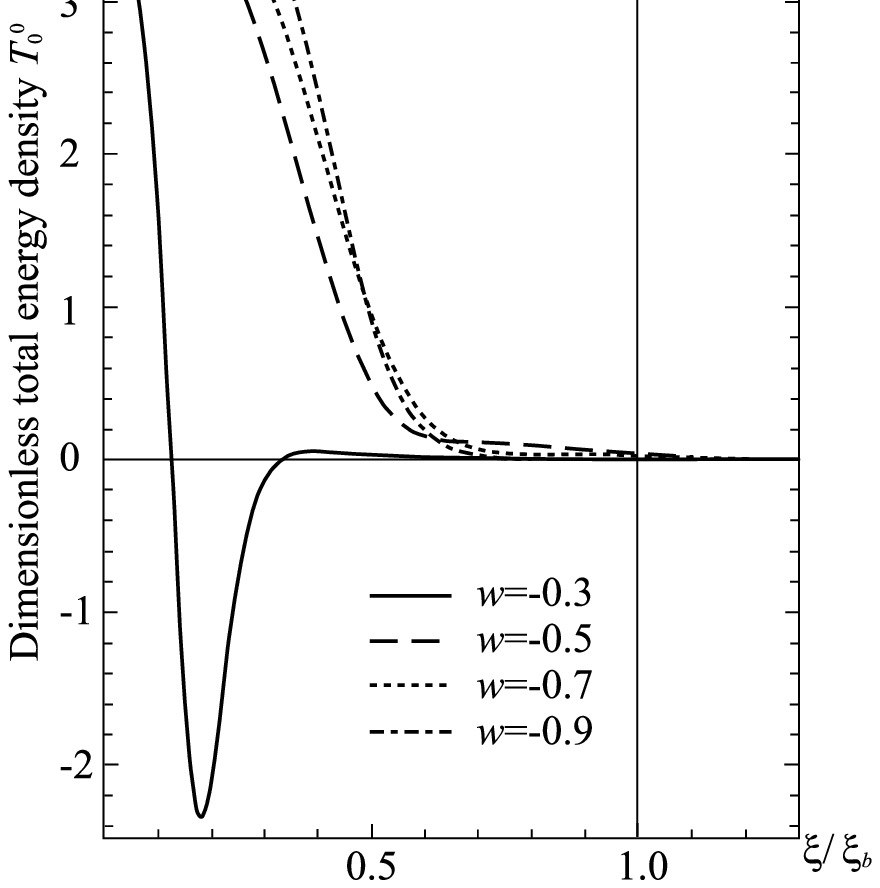}
  \end{center}
\end{minipage}\hfill
\begin{minipage}[t]{.5\linewidth}
  \begin{center}
  \includegraphics[width=7.5cm]{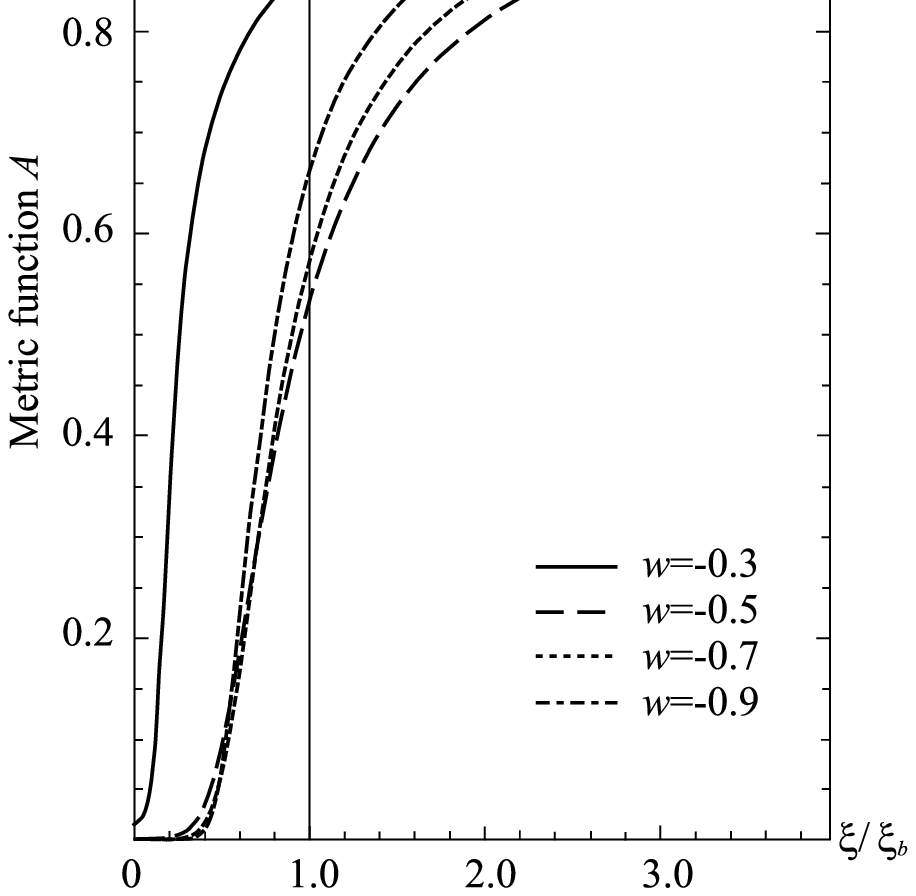}
  \end{center}
\end{minipage}\hfill
\vspace{-1.5cm}
  \caption{Typical distributions of the dimensionless total energy density $T_0^0$
from Eq.~\eqref{tot_energ_dens} (left panel) and graphs of the metric function $A$ (right panel)
are shown as functions of the  relative radius $\xi/\xi_b$
for different values of the equation-of-state parameter~$w$. For both panels,
the central dark matter energy density is taken as $\varepsilon_c=10^{-4}$, and the free parameters appearing in the coupling function
$f$ from Eq.~\eqref{fun_f} are chosen as $a=-150, b=2$.
 The thin vertical line corresponds to the boundary of the dark matter where its pressure and density are equal to zero.
At negative $\xi$, there are corresponding symmetric solutions.}
 \label{energ_metr}
\end{figure}

\begin{figure}[t]
\centering
  \includegraphics[height=7.5cm]{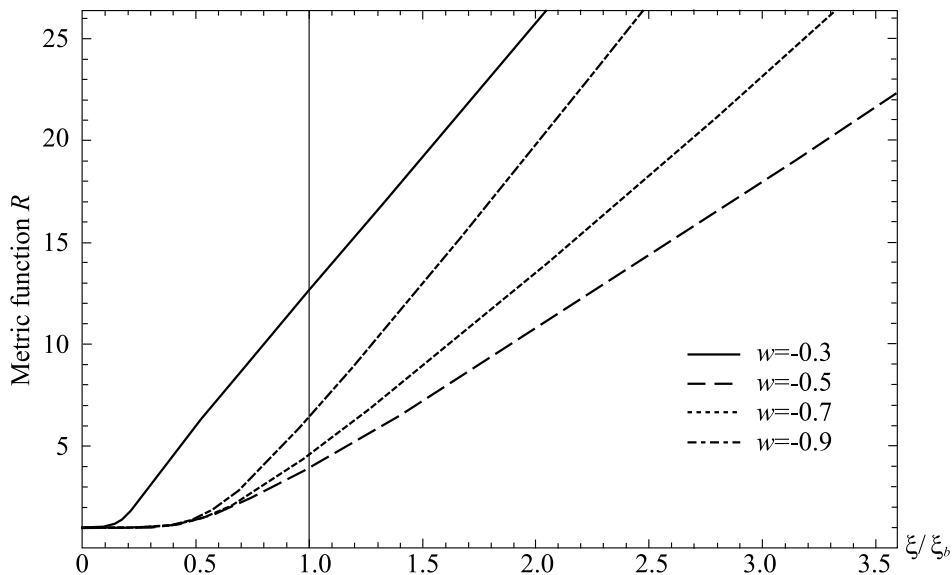}
\caption{The metric function $R$  is shown
as a function of the  relative radius $\xi/\xi_b$ for different values of the equation-of-state parameter
 $w$.
 The values of the parameters $\varepsilon_c, a$, and $b$ are the same as those in Fig.~\ref{energ_metr}.
 At large distances ($\xi/\xi_b \gg 1$) and up to the point where
 the scalar field reaches its cosmological value $\phi_0$, the metric function $R\to \xi$.
}
\label{fig_metr_R}
\end{figure}

Let us now address the question of the total energy density of the configurations under consideration. It is given
by the $(_0^0)$ component of the energy-momentum tensor,
\begin{equation}
\label{tot_energ_dens}
T_0^0=f \varepsilon+\frac{1}{2}  \phi^{\prime 2}+V(\phi)+\frac{\beta}{2}\frac{Q^2}{R^4}\,.
\end{equation}
It contains the contributions from the dark matter energy density $\varepsilon$,
and also from the scalar   and electric/magnetic fields. The numerical calculations show that in considering configurations
with sizes of the order of or less than those of galaxies,  the contribution from the potential energy
$V(\phi)$ is negligibly small on these scales.  The main contribution to the mass comes from
the electric and/or magnetic fields; they determine the size of the throat
[see Eqs.~\eqref{rel_rc_Hc_A} and \eqref{rel_rc_Hc_B}], and correspondingly its mass, and give
significant contributions to the energy density in the internal and external regions of the configuration.
The contributions coming from the terms with the dark matter,
$f \varepsilon$, and the gradient of the scalar field, $\phi^{\prime 2}$, depend on the central energy density
 $\varepsilon_c$ and on the parameter $w$ whose values
determine the fraction of the exotic matter at the center of the
 configuration. The amount of the exotic matter in turn affects directly the central value of the scalar field $\phi_c$,
since the latter is obtained by equating the expression for
 $f$ from Eq.~\eqref{fun_f} at $\phi=\phi_c$ to the expression for $f_c$ from
Eq.~\eqref{expr_R_c_f_c}. Then, starting from $\phi_c$ thus obtained, we seek a solution for $\phi$ which
goes to the background cosmological value $\phi_0$ somewhere at $\xi \gg \xi_b$. In this case
the numerical calculations indicate that, for the values of the parameters used in this paper,
at the point where $\phi$ becomes equal to $\phi_0$, the scalar-field energy density
 $V(\phi_0)$ is always much larger than the energy density of the electric and/or magnetic fields
$[(\beta/2) (Q^2/R^4)]_{\phi=\phi_0}$.
This allows one to regard the configurations in question,
to a good approximation, as embedded only in an external, homogeneous scalar field $\phi_0$.
Notice that if we formally extend the solution
beyond the point
$\phi=\phi_0$, putting here $V(\phi_0)$ equal to a positive constant corresponding to the background energy density
of dark energy (the effective $\Lambda$ term), then at some point the metric function
$A$ inevitably goes to 0 which is a consequence of the fact that these solutions belong to the
Reissner-Nordstr\"{o}m-de Sitter-type solutions.

\begin{figure}[t]
\begin{minipage}[t]{.5\linewidth}
  \begin{center}
  \includegraphics[width=7.5cm]{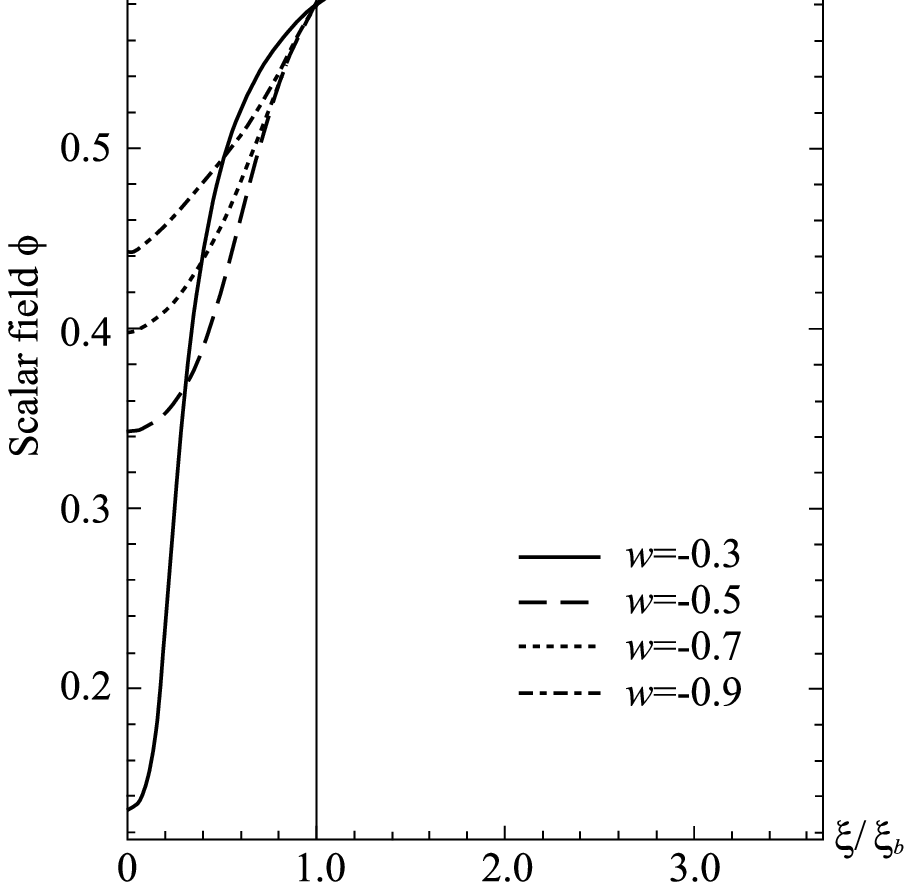}
  \end{center}
\end{minipage}\hfill
\begin{minipage}[t]{.5\linewidth}
  \begin{center}
  \includegraphics[width=7.5cm]{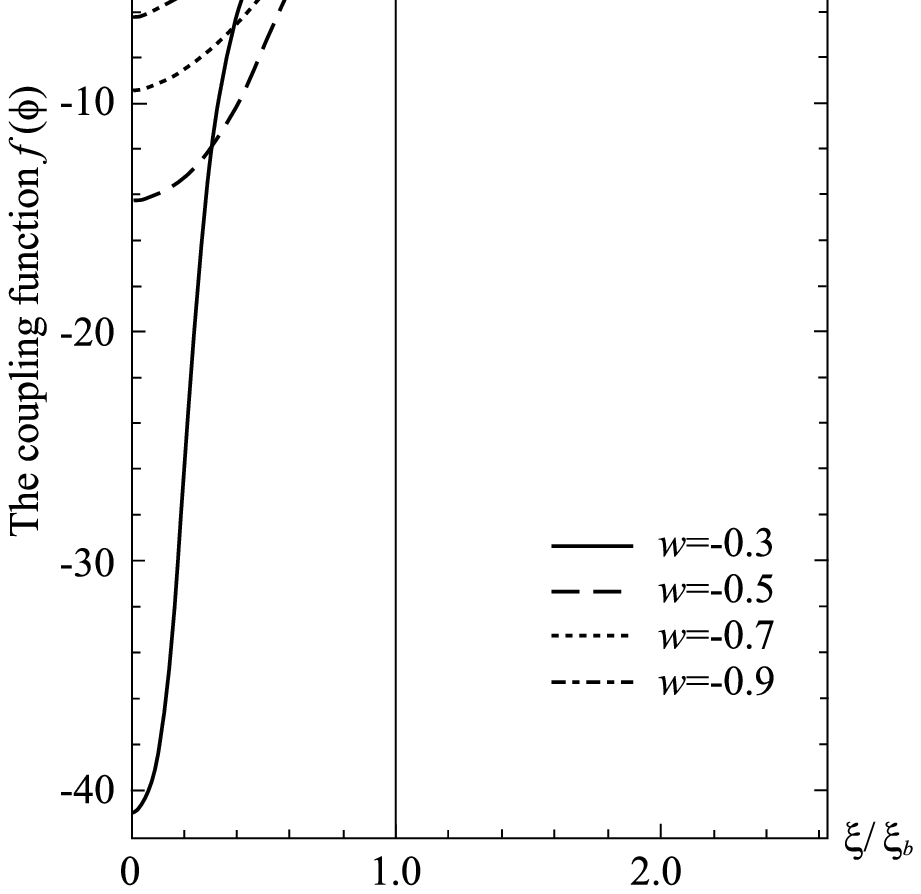}
  \end{center}
\end{minipage}\hfill
  \caption{The scalar field $\phi$  (left panel) and the coupling function $f(\phi)$ (right panel) are shown
as functions of the  relative radius $\xi/\xi_b$ for different values of the equation-of-state parameter
 $w$.
 The values of the parameters $\varepsilon_c, a$, and $b$ are the same as those in Fig.~\ref{energ_metr}.
 At large distances ($\xi/\xi_b \gg 1$), the scalar field goes to the cosmological value $\phi_0$, and the coupling function $f\to 1$.}
 \label{sf_coup}
\end{figure}

Figures~\ref{energ_metr}-\ref{sf_coup} show the results of numerical calculations.
Since the solutions under consideration demonstrate a de Sitter-like behavior,
the metric function $A$ reaches a maximum,
and then drops to zero at some finite value of $\xi$.
By choosing appropriate values of $A_c$, which merely corresponds to a redefinition of
the time coordinate $x^0$ in the metric~\eqref{metric_wh}, the function $A$ is normalized to its maximum value.

As an example, for all graphs in Figs.~\ref{energ_metr}-\ref{sf_coup}, the dimensionless central value of the
dark matter energy density is taken as  $\varepsilon_c=10^{-4}$, and the throat size is
 $R_c=1$ [see Eq.~\eqref{expr_R_c_f_c}].
By choosing some characteristic  value of the mass of a dark matter particle, say $m=1~\text{GeV}$,
and taking into account the expressions for the dimensionless variables  \eqref{dimless_var} used here,
the corresponding dimensional quantities can be presented in the form
\begin{equation}
\label{r_th_rho_c_dms}
r_c=2.41\times 10^5\, \text{cm} \left(\frac{1~\text{GeV}}{m}\right)^2, \quad
\rho_c=2.3\times 10^{17}\frac{\text{g}}{\text{cm}^3}\left(\frac{m}{1~\text{GeV}}\right)^4 \varepsilon_c.
\end{equation}
Here $\rho_c$ is the central mass density of the dark matter.
Since at the moment  it is not definitely known  which type of particles
dark matter consists of,
various particles with different masses are considered in the literature.
This could
be both  superlight gravitinos with a mass of the order of
$10^{-2}~\text{eV}$ and superheavy  weakly interacting massive particles
with a TeV mass scale~\cite{Bertone:2004pz}.
Correspondingly, the dark matter central density and the size of the throat will vary considerably depending on the mass
of particles. The numerical calculations thus show that, for example, if
we assume for definiteness that
$m$ lies in the range $1~\text{eV} \lesssim m \lesssim 10^{2}~\text{GeV}$ used in describing dark matter
particles in the form of fermions \cite{Narain:2006kx}, then the solutions presented in Figs.~\ref{energ_metr}-\ref{sf_coup}
remain practically unchanged. That is, the structure of the configurations under consideration is in essence
independent of the mass of dark matter particles, and the dimensionless solutions obtained enable one
to describe objects whose physical characteristics, in dimensional units, can be found by a simple rescaling of the variables
by using the appropriate dimensional factors from Eq.~\eqref{dimless_var}.

Comparing the expression for $r_c$ from Eq.~\eqref{r_th_rho_c_dms} with Eq.~\eqref{rel_rc_Hc_B},
it is seen that the intensity of the electric and/or magnetic fields at the throat, which is inversely proportional to $r_c$,
decreases with decreasing the mass of dark matter particles.
Then, for instance, at $m\sim 1~\text{GeV}$ we have very compact
objects with sizes and masses
comparable to those of neutron stars, and with the extremely high intensity of the electric and/or magnetic fields at the throat
 $H_c\sim 10^{19}~\text{Gs}$.
When $m\sim 10~\text{keV}$, the throat size is comparable to that of
quasars $\sim 10^{15} \text{cm}$, with the corresponding mass
 $\sim 10^{10} M_\odot$ and $H_c\sim 10^{9}~\text{Gs}$ (cf. the configurations of Ref.~\cite{Kardashev:2006nj}).

It turns out that for
the parameters of the scalar field used here [see  Eq.~\eqref{fun_V} and the subsequent paragraph],
the physical characteristics of
the configurations in question are similar to those of magnetic wormholes from Ref.~\cite{Kardashev:2006nj}.
The numerical calculations indicate that
in all cases considered here, the boundary of the dark matter fluid is situated near
$10\, r_c$. Inside this region, the contribution coming from the exotic matter may be important. In particular,
this results in the appearance of negative energy densities in the inner regions of the configuration
(see Fig.~\ref{energ_metr}, the case with $w=-0.3$). This allows one to reduce the mass of matter in the inner region
by choosing the free parameters of the system in such a way as to provide the presence of a larger amount of the
exotic matter. The possibility of making such a choice is due to the form of the quintessence potential $V(\phi)$
which, being  shallow enough in the neighborhood of its minimum, permits one to change the value of the background field
$\phi_0$ within quite wide limits while maintaining the necessary background energy density
 $V(\phi_0) \simeq 10^{-47}~\text{GeV}^{4}$. This in turn allows one to adjust the central values of the scalar field
 so as  to provide a large amount of the exotic matter in the inner regions of the configuration.

\section{Conclusion}
\label{conclusion}

Starting from the assumption that dark matter may be nonminimally coupled to dark energy,
we have studied spherically symmetric systems with nontrivial topology which is provided by a
proper choice of the nonminimal coupling function $f(\phi)$.
As an explicit example, we have considered the particular case  where dark energy is modeled by a usual
(nonghost) scalar field $\phi$, and dark matter is taken to be a perfect fluid,
which satisfies the null energy condition, and is described by the linear equation of state
$p=w \varepsilon$. For this case we assume that the function  $f(\phi)$ may become negative
at high matter densities which characterize the inner
regions of compact objects,
thereby providing conditions for the violation of the NEC in the vicinity
of the center. Then, by considering some general conditions under which a nontrivial topology might occur
in such a system,
we have shown that
solutions describing traversable wormholes can be obtained only when
1) an electric/magnetic charge is present, and  2) the equation-of-state parameter $w$ is negative.

When these conditions are satisfied, it is possible to construct solutions which may describe configurations
embedded in an external quintessence scalar field. In this case we sought solutions that satisfy the
requirements (i)-(iv) (see the beginning of Sec.~\ref{statem_prob}), and
showed the following.
\vspace{-0.2cm}
\begin{enumerate}
\itemsep=-0.2pt
\item[(1)] 
There exist regular static, spherically symmetric solutions
found numerically by solving the coupled Einstein-matter
equations subject to a set of appropriate boundary conditions.
The obtained solutions describe compact
mixed dark matter/dark energy configurations consisting of two parts: the internal region from the center up to
the radius $\xi=\xi_b$ corresponding to the boundary of the dark matter, where the pressure and density of the dark matter vanish, and
the external region ($\xi > \xi_b$) where there remain only the scalar and electric/magnetic fields.
\item[(2)] Assuming that the effects of the nonminimal coupling are only important at relatively high densities of dark matter,
we sought solutions for the scalar field that started
from some central value $\phi_c$ and went to the cosmological background value $\phi_0$
at $\xi\gg \xi_b$. In doing so, we assumed
that the nonminimal coupling function $f(\phi)$ from Eq.~\eqref{fun_f} is always negative up to the boundary of the
dark matter fluid where it crosses zero, and then
tends to unity as $\phi \to \phi_0$. In this case the free parameters of the system were adjusted so that at this point
the gradient of the scalar field $\phi^\prime_{\phi=\phi_0}=0$.
This enables the solution for the scalar field to be matched smoothly onto the
homogeneous solution $\phi=\phi_0$.
Such configurations may be regarded as embedded
in an external, homogeneous cosmological scalar field
whose energy density is close to the critical one $\simeq 10^{-47}~\text{GeV}^{4}$.
In other words,
the configurations in question may be thought of as embedded in the universe described
by the quintessence  Lagrangian \eqref{lagr_SFDMEF} with $f=1$.
\item[(3)]  On the scales under consideration, the potential energy $V(\phi)$
has no substantial influence on the characteristics of the objects under investigation.  This allows us to suppose
that the use of other quintessence potentials compatible with  observations, instead of that given by
Eq.~\eqref{fun_V}, will give rise to configurations having similar physical parameters.
\end{enumerate}
\vspace{-0.2cm}

The configurations considered in the present paper resemble magnetic wormholes from
Ref.~\cite{Kardashev:2006nj}, in which a magnetic field plays a key role, and exotic matter can contribute
only a small amount.
Such exotic matter can be presented as a massless
 ghost scalar field \cite{Bronnikov:1973fh,Ellis:1973yv,ArmendarizPicon:2002km},
 or, equivalently, as dust matter with negative energy density \cite{Novikov:2007zz}
falling off quite slowly with distance (an inverse quartic dependence). A distinctive feature of the configurations studied here is that
 there is a possibility to concentrate the exotic matter around the throat at distances $\sim 10$ throat radii,
and beyond these limits the ``tails'' of the scalar and electric/magnetic fields, satisfying the null energy condition,
are only present.

\section*{Acknowledgements}

We gratefully acknowledge support provided by the Volkswagen Foundation.
This work was partially supported by the Grant No.~378 in fundamental research in natural sciences
by the Ministry of Education and Science of Kazakhstan.
We also would like to thank the Carl von Ossietzky University of Oldenburg for hospitality while this work was carried out.

\end{document}